# Electrical and magnetic properties of the new Kondo-lattice compound $Ce_3Pd_4Ge_4$

H. J. Im[a] and Y. S. Kwon[a, b,] *

[a] *BK21 Physics Research Division and Institute of Basic Science, SungKyunKwan University, Suwon 440-746, South Korea*

[b] *Center for Strongly Correlated Material Research, Seoul National University, Seuol 151-742, South Korea*

M. H. Jung

*National High Magnetic Field Laboratory, Los Alamos National Laboratory, Los Alamos, NM 87545*

We have measured the electric resistivity, magnetoresistance, magnetic susceptibility and magnetization of the new Kondo-lattice compound $Ce_3Pd_4Ge_4$. The electrical resistivity exhibits a rapid drop at temperatures below 6 K, while the magnetic susceptibility does not show any corresponding anomaly at that temperature. This phenomenon is similar to that of $Ce_3Pd_{20}Ge_6$ which shows quadrupolar interation. We suggest that there is the possibility of quadrupolar interaction in the orthorhombic 4*f*–electron system $Ce_3Pd_4Ge_4$. In addition, it is realized that the spin-dependent scattering effect is responsible for the magnetotransport.

PACS: 71. 20. Eh; 72. 15. Qm

*To whom correspondence should be addressed. E-mail: yskwon@skku.ac.kr

The Ce-Pd-Ge system synthesizing at least 17 ternary compounds [1] is of interest because of their various ground states governed by the competition between the RKKY and Kondo interactions. For instance, antiferromagnetic ground states were observed for dense Kondo compounds, CePdGe ($T_N$ = 3.4 K) [2], CePd$_2$Ge$_2$ ($T_N$ = 5.1 K) [3], Ce$_2$Pd$_3$Ge$_5$ ($T_N$ = 3.8 K) [4] and CePd$_5$Ge$_3$ ($T_N$ = 1.9 K) [5]. It was reported that Ce$_3$Pd$_{26}$Ge$_6$ ($T_N$ = 0.7 K) shows anisotropic magnetic phase diagram with quadrupolar ordering [6]. Although Ce$_3$Pd$_4$Ge$_4$ is known to be isostructural to Ce$_3$Ni$_4$Ge$_4$ (orthorhombic space group of Immm) [1, 7], its physical properties have not been reported yet. This paper will introduce the electrical and magnetic properties of Ce$_3$Pd$_4$Ge$_4$ by measurements of electrical resistivity, magnetoresistance, magnetic susceptibility and magnetization.

The polycrystalline samples of Ce$_3$Pd$_4$Ge$_4$ were prepared by arc melting the constituent elements in an argon atmosphere, followed by annealing at 800 $^o$C in a vacuum for one week. The structural characterization by X-ray powder diffraction agrees with that reported in Ref. 1. The electrical resistivity was measured by a standard four-probe AC method. The magnetization measurements were performed using a Quantum Design PPMS system.

Figure 1 shows the inverse magnetic susceptibility $\chi^{-1}(T)$ measured in a field of 2 kOe. Above 50 K, $\chi^{-1}(T)$ follows a Curie-Weiss law, giving the effective magnetic moment $\mu_{eff}$ = 2.57 $\mu_B$ and the paramagnetic Curie temperature $\theta_P$ = −33.17 K. The $\mu_{eff}$ value is close to that expected for a trivalent Ce ion and the $\theta_P$ value is rather large, similar to those found in many Ce-based Kondo compounds. The deviation from the Curie-Weiss behavior at low temperature is possibly due to the crystal field effect. Down to 2 K, $\chi(T)$ does not show any anomaly, which could hint on long-range magnetic ordering. The inset of Fig. 1 shows the magnetization $M(H)$ curve measured at 2 K. With increasing magnetic field, $M(H)$ exhibits a upward curvature, which may be a precursor to magnetic order at low temperatures, and is linear up to 7 T. A very weak hysteresis loop without remanence was observed around 3 T.

The electrical resistivity $\rho(T)$ is shown in Fig. 2. As the temperature is lowered, $\rho(T)$ monotonically decreases down to about 20 K and then rapidly decreases below $T = 6$ K. Since $\chi(T)$ shows no appreciable anomaly at 6 K, the nature of this anomaly cannot be attributed to long-range magnetic order as noted above. It will be discussed in relation to the possibility of multipolar (e.g. quadrupolar) moment or lattice instability. In the inset of Fig. 2, the low-temperature data of $\rho(T)$ can be expressed as $\rho_o + AT^2$ indicating Kondo lattice behavior. The values of $\rho_o$ and A are estimated to be 2.6 μcm and 0.82 μcm/K$^2$ (2 K ≤ T ≤ 3.5 K) and 6.5 μcm and 0.53 μcm/K$^2$ (3.7 K ≤ T ≤ 5 K), respectively. The A values are much larger than typical transition metals [8] but smaller than dense Kondo compounds [9].

If the ground state is $\Gamma_8$ quartet under a crystalline field of cubic symmetry, there exist both RKKY and quadrupolar interactions. The RKKY interaction between Ce ions leads to a magnetic order and the quadrupolar interaction often leads to a qudrupolar ordering or a lattice instability. Such effects manifest themselves at low temperatures in cubic systems, CeAg, [10] CeB$_6$ [11] and Ce$_3$Pd$_{20}$Ge$_6$ [12], where both quadrupolar and magnetic orderings were found as closely correlated. These anomalies observed in CeAg, CeB$_6$ and Ce$_3$Pd$_{20}$Ge$_6$ resemble the case of orthorhombic system Ce$_3$Pd$_4$Ge$_4$, which shows an anomaly at 6 K in $\rho(T)$ but no corresponding anomaly at 6 K in $\chi(T)$. Although the crystalline-field ground state of $Ce_3Pd_4Ge_4$ has not been known yet, we are tempted to assume that the anomaly of $\rho(T)$ at 6 K is real, probably due to a lattice instability caused by a multipolar ordering. If it is proved, $Ce_3Pd_4Ge_4$ would be the first compound with an orthorhombic crystal structure revealing a multipolar ordering.

Figure 3 shows the normalized magnetoresistance $\Delta\rho/\rho_o$ measured at low temperatures in both longitudinal ($H//I$) and transverse ($H\perp I$) configurations. At 2.2 K, $\Delta\rho/\rho_o$ for $H//I$ is initially negative and then turns positive at 1.2 T, followed by a broad peak at 9 T. The field range of negative $\Delta\rho/\rho_o$ corresponds to that $M(H)$ is rapidly enhanced, and the range of positive $\Delta\rho/\rho_o$ corresponds to that $M(H)$ is relatively slowly increased. Thus, the slope change of $\Delta\rho/\rho_o$ can be interpreted in light of the spin-dependent scattering effect. This field dependence is quite similar to that for $H\perp I$, where $\Delta\rho/\rho_o$ shows an upturn at 1.4 T and makes a broad peak at 12 T. With increasing temperature, the broad peak shifts to lower fields, then $\Delta\rho/\rho_o$ at 6 K becomes negative over all the field range measured up to 18 T. The negative values of $\Delta\rho/\rho_o$ are −46% and −30% for $H//I$ and $H\perp I$, respectively.

This work was supported by the Korea Science and Engineering Foundation through the Center for Strongly Correlated Materials Research at Seoul National University and by the Korea Research Foundation Grant (KRF-2001-DP0139).

**Figure captions**

Fig. 1. Inverse magnetic susceptibility $\chi^{-1}(T)$ measured in a field of 2 kOe. The inset shows the $M(H)$ curve at 2 K.

Fig. 2. Electrical resistivity $\rho(T)$ in the logarithmic temperature scale. The inset shows the low-temperature data expressed as $\rho(T) = \rho_o + AT^2$.

Fig. 3. Normalized magnetoresistance $\Delta\rho/\rho_o$ measured at 2.2 and 6 K in the longitudinal ($H//I$) and transverse ($H \perp I$) configurations.

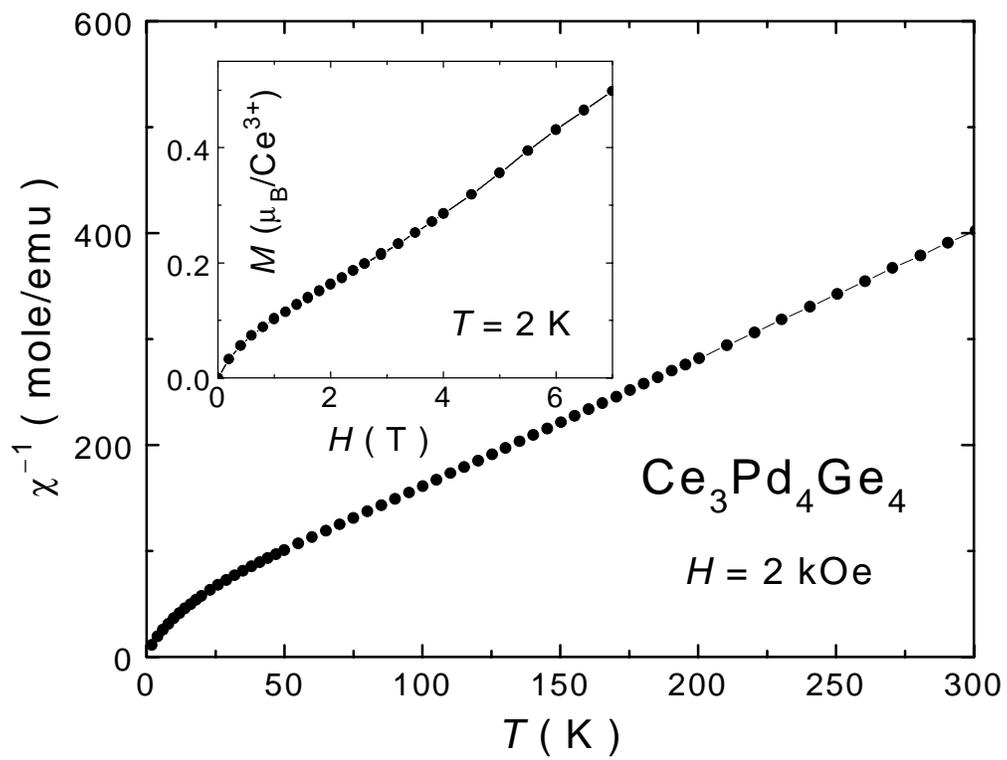

Fig. 1. H. J. Im et al.

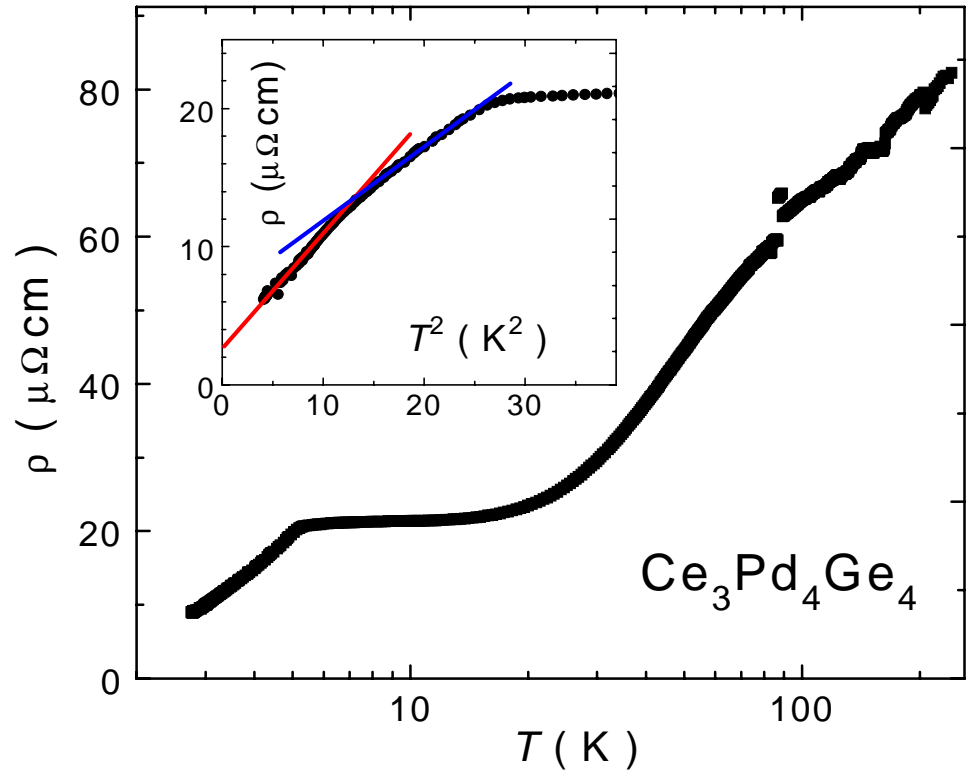

Fig. 2. H. J. Im et al.

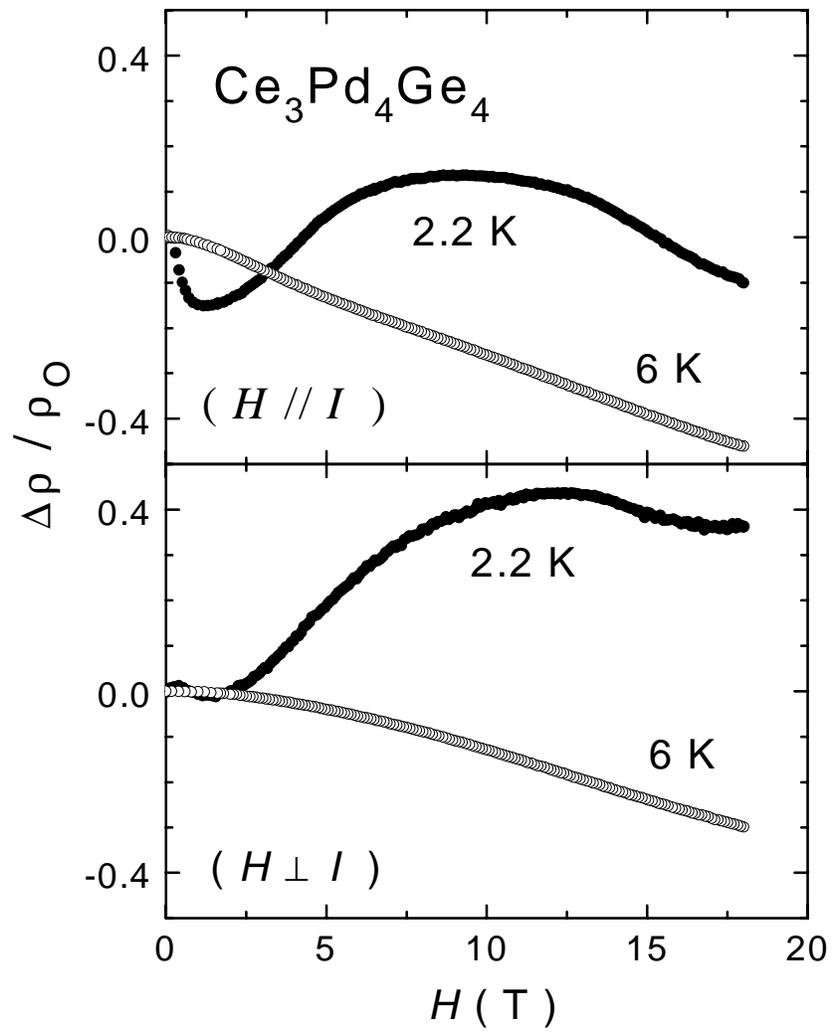

Fig. 3. H. J. Im et al.